\documentclass{icrc2009}

\usepackage{graphicx}   % for including figures
\usepackage[font=footnotesize]{subfig} % subfig.sty for a double column floating figure using two subfigures
\usepackage{fixltx2e}
\usepackage{amsmath}
\usepackage{amssymb}
\usepackage{url}

\newcommand{\shorttitle}[1]%
{\markboth{Proceedings of the 31\MakeLowercase{$^{st}$} ICRC, {\L}\'{o}d\'{z} 2009}{#1} }
\newcommand{\etal}{\MakeLowercase{\textit{et al. }}} % "et al."

\begin{document}
\title{Elemental energy spectra of cosmic rays  measured by CREAM-II}

\author{\IEEEauthorblockN{
P. Maestro\IEEEauthorrefmark{1},
H.~S. Ahn\IEEEauthorrefmark{2},
P. Allison\IEEEauthorrefmark{4}, 
M.~G. Bagliesi\IEEEauthorrefmark{1}, 
L. Barbier\IEEEauthorrefmark{5},
J.~J. Beatty\IEEEauthorrefmark{4}, 
G. Bigongiari\IEEEauthorrefmark{1}, \\
T.~J. Brandt\IEEEauthorrefmark{4}, 
J.~T. Childers\IEEEauthorrefmark{6}, 
N.~B. Conklin\IEEEauthorrefmark{7},
S. Coutu\IEEEauthorrefmark{7}, 
M.~A. DuVernois\IEEEauthorrefmark{6},
O. Ganel\IEEEauthorrefmark{2}, 
J.~H. Han\IEEEauthorrefmark{2}, \\
J.~A. Jeon\IEEEauthorrefmark{8}, 
K.~C. Kim\IEEEauthorrefmark{2},
M.~H. Lee\IEEEauthorrefmark{2}, 
A. Malinine\IEEEauthorrefmark{2}, 
P.~S. Marrocchesi\IEEEauthorrefmark{1}, 
S. Minnick\IEEEauthorrefmark{9}, 
S.~I. Mognet\IEEEauthorrefmark{7}, \\
S.~W. Nam\IEEEauthorrefmark{8}, 
S. Nutter\IEEEauthorrefmark{10},
I.~H. Park\IEEEauthorrefmark{8}, 
N.~H. Park\IEEEauthorrefmark{8}, 
E.~S. Seo\IEEEauthorrefmark{2}\IEEEauthorrefmark{3}, 
R. Sina\IEEEauthorrefmark{2}, 
P. Walpole\IEEEauthorrefmark{2}, 
J. Wu\IEEEauthorrefmark{2}, \\
J. Yang\IEEEauthorrefmark{8}, 
Y.~S. Yoon\IEEEauthorrefmark{2}\IEEEauthorrefmark{3}, 
R. Zei\IEEEauthorrefmark{1} and  
S.~Y. Zinn\IEEEauthorrefmark{2}}

     \\
\IEEEauthorblockA{\IEEEauthorrefmark{1}Department of Physics, University of Siena and INFN, Via Roma 56, 53100 Siena, Italy}
\IEEEauthorblockA{\IEEEauthorrefmark{2}Institute for Physical Science and Technology, University of Maryland, College Park, MD 20742, USA}
\IEEEauthorblockA{\IEEEauthorrefmark{3}Department of Physics, University of Maryland, College Park, MD 20742, USA}
\IEEEauthorblockA{\IEEEauthorrefmark{4}Department of Physics, Ohio State University, Columbus, OH 43210, USA}
\IEEEauthorblockA{\IEEEauthorrefmark{5}Astroparticle Physics Laboratory, NASA Goddard Space Flight Center, Greenbelt, MD 20771, USA} 
\IEEEauthorblockA{\IEEEauthorrefmark{6}School of Physics and Astronomy, University of Minnesota, Minneapolis, MN 55455, USA}
\IEEEauthorblockA{\IEEEauthorrefmark{7}Department of Physics, Penn State University, University Park, PA 16802, USA}
\IEEEauthorblockA{\IEEEauthorrefmark{8}Department of Physics, Ewha Womans University, Seoul 120-750, Republic of Korea}
\IEEEauthorblockA{\IEEEauthorrefmark{9}Department of Physics, Kent State University, Tuscarawas, New Philadelphia, OH 44663, USA}
\IEEEauthorblockA{\IEEEauthorrefmark{10}Department of Physics and Geology, Northern Kentucky University, Highland Heights, KY 41099, USA} 
}

% please write the preseter's name and short title (3-4 words maximum)
%    which will appear at the header of the even pages.
\shorttitle{P. Maestro \etal Cosmic-ray spectra with CREAM-II}
\maketitle

\begin{abstract}
We present new measurements of the energy spectra of cosmic-ray (CR) nuclei
from the second flight of 
the balloon-borne experiment CREAM
(Cosmic Ray Energetics And Mass). 
%The CR charge identification 
%was based on different techniques (Cerenkov light, specific ionization in scintillators and
% silicon sensors), while the particle energy was 
% measured by a thin ionization calorimeter.
The instrument (CREAM-II) was comprised of 
detectors based on different techniques (Cherenkov light, specific ionization in scintillators and
 silicon sensors) to provide a redundant charge identification and 
a thin ionization calorimeter capable of measuring the energy of cosmic rays 
up to several hundreds of TeV.
The data analysis is described and   
the individual energy spectra of C, O, Ne, Mg, Si and Fe
are reported up to $\sim 10^{14}$ eV. The spectral shape looks
nearly the same for all the primary elements and can be expressed as 
a power law in energy $E^{-2.66 \pm 0.04}$.
The nitrogen absolute intensity 
in the energy range 100-800 GeV/n is also measured. 
\end{abstract}

 
\section{Introduction}
CREAM is a balloon-borne experiment designed to directly measure 
the elemental composition  and  the energy spectra of 
cosmic rays from H to Fe %hydrogen to iron
in the energy range 10$^{11}$-10$^{15}$ eV. 
CREAM aims to experimentally test %The main physics goal addressed by CREAM is 
%the experimental test of the validity of 
astrophysical models proposed to 
explain the acceleration mechanism 
of cosmic rays and their propagation in the Galaxy \cite{1}.\\
Since 2004, four instruments were  successfully flown on long-duration balloons
in Antarctica. 
The instrument configurations varied slightly in each mission,
due to various detector upgrades.
In this paper, we describe the procedure used to analyze  the data
collected during the second flight and reconstruct the energy spectra 
of the major primary CR nuclei and nitrogen. Results are compared with other
direct observations and discussed.\\
\section{The CREAM-II instrument}
The instrument for the second flight included: a
redundant system for particle identification, consisting 
(from top to bottom)
of a timing-charge detector (TCD), a
Cherenkov detector (CD), a pixelated silicon charge detector (SCD), and  
a sampling imaging calorimeter (CAL) designed to provide a
measurement of the energy of primary nuclei in the multi-TeV region. \\
The TCD  is comprised of two planes
of four 5 mm-thick plastic scintillator paddles, 
read out by fast  photomultiplier tubes (PMTs) and covering an area of 120$\times$120 cm$^2$.
%A hodoscope of squared scintillating fibers (S3) located above the calorimeter
%provides a reference time for TCD.
It was designed to determine each element charge  
with a resolution $\lesssim$ 0.35 $e$  (units of the electron charge) and
discriminate against albedo particles \cite{TCD}. 
The CD is a 1 cm-thick plastic radiator, with 1 m$^2$ surface area, read out by eight PMTs via wavelength shifting bars. 
It is mainly used to flag relativistic particles.
The SCD   is a dual layer of 312 silicon sensors, each segmented as an array of 4$\times$4 
pixels which covers an effective area of 0.52 m$^2$ with no dead regions.
It is capable of resolving individual  elements  
from H to Ni with a fine charge resolution \cite{SCD}.\\
The CAL  is a stack of 20 tungsten plates (50$\times$50 cm$^{2}$ and 1 radiation-length thick)
  interleaved with active layers of
%instrumented with 
1 cm-wide scintillating fiber ribbons, read out by 40 hybrid photodiodes  \cite{CAL}. % (HPD).
%The surface area of the detector is 50$\times$50 cm$^{2}$ 
%and its thickness is about 9 cm and corresponds to 20 radiation lengths.
A 0.47 $\lambda_{int}$-thick graphite target precedes the CAL, and serves
to induce hadronic interactions of the CR nuclei.
The electromagnetic (e.m.) core of the resulting hadronic cascade is imaged by the CAL which is  sufficiently thick   
to contain the shower maximum and grained finely enough to reconstruct the direction of the incident particle. \\
CREAM-II  was launched on Dec. 16$^{th}$  2005 from McMurdo  and floated
over Antarctica for 28 days, %until Jan. 13$^{th}$ 2006,  at a
at a balloon  altitude  of 35 to 40 km.
All the detectors performed well in flight.
The data acquisition was enabled by the trigger system 
whenever a shower developing in at least 6 planes was detected in the CAL or
a relativistic cosmic ray with Z $\geq$ 2 was identified by the TCD and CD. 
%The trigger system kept both the showers developing in at least 6 CAL planes
%and any cosmic ray identified by the TCD and CD as relativistic with  Z $\geq$ 2.
A total of 57 GB of data were collected.
A detailed description of the instrument and its performance in flight is given in \cite{Marrocchesi}.\\
\section{Data analysis}
The present analysis uses a subset of data 
 collected in the period Dec. 19$^{th}$- Jan. 12$^{th}$, having stable instrument conditions.
In order to measure the elemental energy spectra, each detected CR particle
has been assigned a charge and an energy. 
An accurate reconstruction of the particle trajectory through the instrument is needed
for the charge determination.
\subsection{Trajectory reconstruction}
The axis of the shower imaged by the calorimeter can be measured to determine 
the arrival direction of a CR nucleus. 
Candidate track points are sampled along the longitudinal development of the shower.
They are defined as the center of gravity of the cluster
formed in each CAL plane by the cell with the maximum signal and its two neighbours.  
A track is built by matching  the candidate  points 
and the shower axis parameters are calculated with a linear $\chi^2$ fit of the track.
Track quality is assessed by requiring a value of  $\chi^2<\,$10.
The reconstructed shower axis is back-projected 
through the graphite target to its intersections with the two  SCD layers. These
define the impact points
of the primary particle, with a spatial resolution better than 1 cm rms.
\subsection{Charge assignment}
In each SCD layer, the pixel with the highest signal is sought
inside a fiducial circle with a 3 cm radius 
centered on the impact point. 
In order to achieve high purity samples of CR elements, 
the signals of the two selected pixels %$S_{bottom}$ and $S_{top}$, 
are  required to be consistent within 30\%. 
%If this cut is satisfied, 
In such a case, the
two independent samples of the specific ionization $dE/dx$ 
are then corrected for the pathlength (estimated from the track parameters)
%($\propto 1/\cos{\left(\theta\right)}$, where $\theta$ is the angle between the particle trajectory and the axis of the instrument) 
traversed by the particle in the silicon sensors. 
The accuracy of the pathlength correction is improved by  
adding the matched SCD pixels to the track points and refitting them again.
By combining the two corrected measures of $dE/dx$,
%of the specific ionization, 
the primary particle charge $Z$ is estimated.
%%%%%%%%%%%%%%%%%%%%%%%%%%%%%%%%%%%%%%%%%%%%%%%%%%%%%%%%%%%%%%%%%%%%%%%%%%%%%%%%%%%%%%%%%
\begin{figure}[!t]
\centering
\includegraphics[scale=0.4]{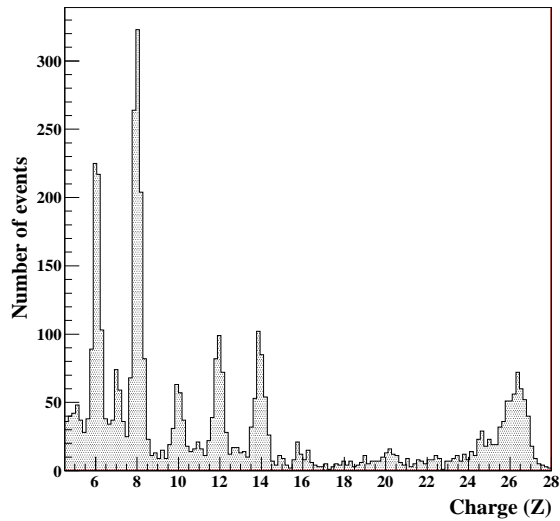}
\caption{Charge histogram obtained by the SCD in the elemental range from B to Fe.
The distribution is only indicative of the SCD charge resolution and the relative elemental abundances are not
meaningful.}
\label{fig1}
\end{figure}
%%%%%%%%%%%%%%%%%%%%%%%%%%%%%%%%%%%%%%%%%%%%%%%%%%%%%%%%%%%%%%%%%%%%%%%%%%%%%%%%%%%%%%%%%
The charge distribution reconstructed by the SCD is shown in Fig.~\ref{fig1};
by fitting each peak to a gaussian, a charge resolution 
$\sigma$ is estimated as:
0.2 $e$ for C, N, O; $\sim$ 0.23 $e$ for Ne, Mg, Si; $\sim$ 0.5 $e$ for Fe.
Samples of each element up to Si are selected with a 
 2$\sigma$ cut around the mean charge value, while for iron a
1$\sigma$ cut is used.
In this way, 1288 O are retained, as well as 456 Si 
and 409 Fe candidates. 
%\vspace{-0.3cm}
\subsection{Energy measurement and deconvolution}
The CREAM thin ionization calorimeter 
samples  the e.m.
core of the hadronic cascade 
initiated by a cosmic ray interacting in the target.
According to the results of tests with  particle beams \cite{Ahn}
and Monte Carlo (MC) simulations based on the FLUKA 2006.3b package
\cite{fluka}, though
a significant part of the cascade energy  leaks out of the calorimeter,
the primary particle energy still scales linearly with the energy 
deposited in the CAL by the shower  core 
up to several hundreds of TeV. 
The total  energy deposit 
is measured by summing up the calibrated signals of all the CAL cells. 
Once  each cosmic ray is assigned an energy, the reconstructed particles of each nuclear
species are sorted into energy intervals 
with width %chosen to be 
commensurate with the  rms resolution of the calorimeter.\\ 
Due to the finite energy resolution of the detector, of the order of 30\% for 
heavy nuclei with energy $>$ 1 TeV,
an unfolding procedure has to be applied to correct 
the measured counts  
in each energy interval  for overlap with the neighbouring bins. 
This requires  solving a set of linear equations
\begin{equation*}
M_i = \sum_{j = 1}^{n} A_{ij} N_j \;\;\;\;\;\;\;\; i=1,..., m
\end{equation*}
relating the 
``true'' counts $N_j$ of $n$ incident energy bins 
to the measured counts $M_i$ of $m$ deposited energy bins.
A generic element of the mixing matrix $A_{ij}$ represents 
the probability that a CR particle, carrying an energy 
corresponding to a given energy  bin \emph{j}, produces 
an energy deposit in the calorimeter falling in the bin \emph{i}. 
The unfolding matrix elements are estimated by analyzing 
MC simulated CR events
with the same procedure as that 
used for the flight data. 
\section{Energy spectrum}
The absolute differential intensity $\Phi$ at the top of the atmosphere          
is calculated by dividing the unfolded number of events  $N_i$ by the bin width $\Delta E_i$, according to  the formula
\begin{equation*}
\Phi(\hat{E}_i) = \frac{N_i}{\Delta E_i}\times \frac{1}{\epsilon \times \text{TOI} \times \text{TOA} \times S\Omega \times \text{T} }
\end{equation*}
where T is the exposure time, $\epsilon$ the efficiency of the selection cuts, $S\Omega$ the geometric factor
of the instrument, and TOI and TOA are, respectively, the corrections to the top of instrument
and to the top of the atmosphere. Each bin is 
centered at a median  energy $\hat{E}_i$ calculated as described in  \cite{LW}.\\
The geometric factor $S\Omega$ is estimated to be 0.46 m$^2$sr
from MC simulations.
The selected set of data amounts to a live time T of 
16 days and 19 hours, close to 75\% of the real time of data taking, as measured by the 
housekeeping system onboard.
The MC estimated overall reconstruction efficiency $\epsilon$ %, estimated from MC simulations,
has a constant value of around 70\% at energies $>$
3 TeV for all nuclei.\\
The probability that a nucleus undergoes a spallation reaction in the 
 material  ($\sim$ 4.8 g/cm$^2$) above
the upper SCD plane is also estimated from MC simulations.
The fraction of surviving nuclei, i.e., the TOI correction, 
spans from 81.3\% for C %70.6\% for Si, 
to 61.9\% for Fe.
The TOA correction is calculated  by simulating with FLUKA  the atmospheric
 overburden
during the flight (3.9 g/cm$^2$ on average).
Survival probabilities
ranging from 84.2\% for C to %77.5\% for Si,   
71.6\%  for Fe  are found.\\
The TOI and TOA corrections take into account only the loss of primary particles
interacting in the atmosphere and in the instrument materials. 
Furthermore, 
the counts $N_i$ for each species are corrected for the gain 
of secondary particles produced by spallation reactions
of heavier CR nuclei. These corrections are of the order of few percent for all the considered elements
but nitrogen. In fact, 
$\sim\,$30\% of the observed N nuclei originate from the spallation of much more abundant oxygen.\\
\section{Results}
The differential intensity at the top of the atmosphere as measured by CREAM-II 
for the major primary CR nuclei from carbon to iron
 is plotted as a function of the 
kinetic energy per nucleon in Fig.~\ref{fig2}. 
The nitrogen energy spectrum (multiplied by $E^{2.5}$)  is shown
in Fig.~\ref{fig3}. 
The error bars in the intensity are calculated as the sum in quadrature 
of the error due to the counting statistics and the systematic uncertainties.\\
The main systematic uncertainties 
stem from the reconstruction algorithm and from the TOI and TOA corrections. In the first case, 
the fractional systematic  error is estimated to be of order 10\% in the energy bins below 3 TeV and 5\% above.
The systematic error for the uncertainty in correcting 
the differential intensity to the top of instrument is 2\%
for the primary elements. For nitrogen, a 15\% error is assigned because of the large contamination 
of O nuclei which spallate into N.
Similar values  are estimated for the systematic error deriving from the uncertainties in the atmospheric
secondary corrections.\\
%The uncertainties in the corrections for secondary nuclei 
%produced in the atmosphere and in the instrument contribute 
%a 2\% each
%to the fractional systematic error of the differential intensity for all the elements, but nitrogen.
%In this case, they contribute a 15\% each, 
%because of the large contamination 
%of O nuclei which spallate into N.\\
In summary, the particle energy range measured by CREAM-II 
extends from around 800 GeV up to 100 TeV.
The absolute intensities are presented without any arbitrary normalization to previous data
and cover a range of six decades.\\
\section{Discussion}
%%%%%%%%%%%%%%%%%%%%%%%%%%%%%%%%%%%%%%%%%%%%%%%%%%%%%%%%%%%%%%%%%%%%%%%%%%%%%%%%%%%%%%%%%
\begin{figure*}[th]
\centering
\includegraphics[scale=0.82]{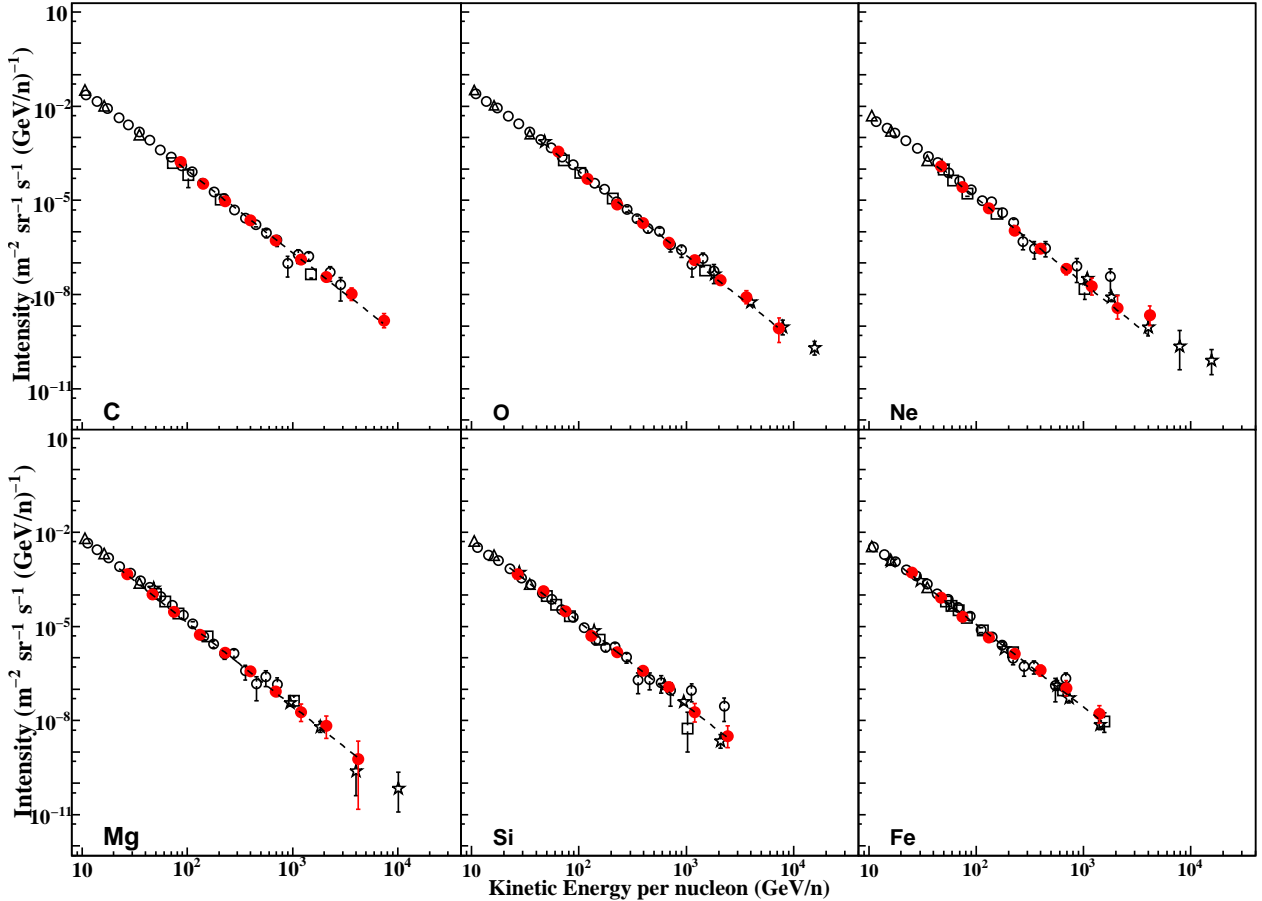}
\caption{Energy spectra of the more abundant heavy nuclei. 
Results of CREAM-II (filled circles)
are compared with measurements from HEAO-3-C2 \cite{HEAO} (triangles), CRN \cite{CRN} (squares), ATIC-2 \cite{ATIC} (open circles), 
 TRACER \cite{TRACER} (stars). The dashed line represents a power-law  fit to the CREAM-II data.
The fitted values of the spectral indices  are, respectively: C $2.61\pm0.07$; O $2.67\pm0.07$; Ne $2.72\pm0.11$; Mg $2.66\pm0.08$; Si $2.67\pm0.08$; Fe $2.63\pm0.12$.
}
\label{fig2}
\end{figure*}
%%%%%%%%%%%%%%%%%%%%%%%%%%%%%%%%%%%%%%%%%%%%%%%%%%%%%%%%%%%%%%%%%%%%%%%%%%%%%%%%%%%%%%%%%
%%%%%%%%%%%%%%%%%%%%%%%%%%%%%%%%%%%%%%%%%%%%%%%%%%%%%%%%%%%%%%%%%%%%%%%%%%%%%%%%%%%%%%%%%
\begin{figure}[h]
\centering
\includegraphics[scale=0.43]{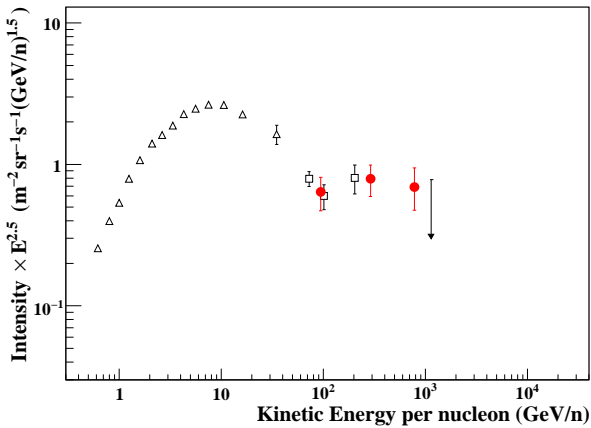}
\caption{Nitrogen energy spectrum multiplied by E$^{2.5}$.
CREAM-II results ({\em filled circles}) 
are compared with previous observations by
HEAO-3-C2 \cite{HEAO}  ({\em triangles}) and CRN \cite{CRN2}  ({\em squares}). 
The arrow is an upper limit set by CRN at 1.1 TeV/n.
}
\label{fig3}
\end{figure}
%%%%%%%%%%%%%%%%%%%%%%%%%%%%%%%%%%%%%%%%%%%%%%%%%%%%%%%%%%%%%%%%%%%%%%%%%%%%%%%%%%%%%%%%%
The CREAM-II results for primary nuclei
are found  in general to be in good agreement with previous measurements of
space-based (HEAO-3-C2 \cite{HEAO}, CRN \cite{CRN}) and 
balloon-borne (ATIC-2 \cite{ATIC}, TRACER \cite{TRACER}) 
 experiments.
A power-law fit $E^{-\gamma}$ to our data indicates 
that the intensities of the more abundant evenly charged heavy elements
have a very similar energy dependence. That is,  their spectra
are characterized, within errors, by nearly
the same spectral index $\gamma$. 
Our observations, based on a calorimetric measurement of the CR energy, 
confirm the results recently reported by the TRACER collaboration, 
using  completely different techniques to determine  the particle energy.
The weighted average of the fitted spectral indices (Fig.~\ref{fig2})
is 
 $\bar{\gamma} = 2.66 \pm 0.04$, consistent, within the error,
with the value of $2.65 \pm 0.05$
obtained from a fit to the combined CRN and TRACER data  \cite{TRACER}.
The great similarity of the spectral indices 
suggests that 
the same mechanism is responsible for the source acceleration  of the primary heavy nuclei,
%primary heavy nuclei are accelerated at the source by the same physical mechanism,
assuming, as common, that the escape pathlength of primary elements from the galaxy does not 
depend on the particle charge for primary elements.\\
The nitrogen data collected by CREAM-II are statistically 
more significant at high energy than any previous observation.
Unlike the primary heavy nuclei, nitrogen is 
mostly produced by spallation in the interstellar medium
but also has  a  primary contribution of order 10\%, as recently measured by CREAM-I \cite{Ahn2}. 
%A simultaneous power-law fit of the HEAO-3-C2, CRN and CREAM-II data sets 
%above 30 GeV/n
%gives a spectral index  of $2.75 \pm 0.15$, which, in spite of the large error, 
%is slightly steeper than the value for primaries, 
%as expected for secondary nuclei.
We notice that, at energies above 100 GeV/n, the nitrogen spectrum
flattens out from the steep decline which characterizes the energy range 10-100 GeV/n.
This supports the hypothesis of the presence of two components in the cosmic nitrogen.
In fact, since the escape pathlength decreases rapidly with energy, 
the secondary nitrogen is expected to become negligible  
at high energy, where only the primary component should remain. This might result in a change of the 
spectral slope, as observed.\\
To provide a check to our measurements, we calculate the
relative abundance ratio of N/O 
and compare it with the CREAM-I result. A N/O ratio 
 (at the top of the atmosphere) equal to $0.080 \pm 0.025 \text{ (stat.)} \pm 0.025 \text{ (sys.)}$ 
is measured 
at an energy of $\sim $800 GeV/n, in good agreement with 
the CREAM-I result \cite{Ahn2}.\\
\section{Conclusion}
The CREAM-II instrument carried out measurements of high Z cosmic ray nuclei
with an excellent charge resolution and  a reliable energy determination.\\
The energy spectra of the major primary heavy nuclei from C to Fe
were measured up to 10$^{14}$ eV and found to agree well with earlier direct measurements.
A new measurement of the nitrogen intensity  in an energy region thus far experimentally unexplored 
indicates a less steep power-law trend in the spectrum with respect to lower energies.
\footnotesize
\section*{Acknowledgments}
This work is supported by NASA, NSF and CSBF in USA, INFN and PNRA in Italy, KICOS and MOST in Korea. 

\end{document}